\documentclass[11pt,a4paper]{article}
\usepackage{amsmath,amsthm,amssymb,epsfig,latexsym}
\newcommand{\be}[1]{\begin{equation}\label{#1}}
\newcommand{\ba}[1]{\begin{eqnarray}\label{#1}}
\newcommand{\ee}{\end{equation}}
\newcommand{\ea}{\end{eqnarray}}
\newcommand{\non}{\nonumber\\\rule{0pt}{30pt}}
\newcommand{\nona}[1]{\nonumber\\\rule{0pt}{#1pt}}

\newcommand{\dis}{\displaystyle}
\newcommand{\eq}[1]{(\ref{#1})}
\newcommand{\tr}{\mathop{\rm tr}}
\newcommand{\Res}{\mathop{\rm Res}}

\newtheorem{thm}{Theorem}[section]
\newtheorem{prop}{Proposition}[section]
\newtheorem{lemma}{Lemma}[section]
\newtheorem{cor}{Corollary}[section]
\newtheorem{Def}{Definition}[section]
\makeatletter
\@addtoreset{equation}{section}
\makeatother

\begin{document}
\begin{flushright}
LPENSL-TH-01/04\\
\end{flushright}
\par \vskip .1in \noindent

\vspace{24pt}

\begin{center}
\begin{LARGE}
{\bf Master equation for spin-spin
correlation functions of the $XXZ$ chain}
\end{LARGE}

\vspace{50pt}

\begin{large}

{\bf N.~Kitanine}\footnote[1]{LPTM, UMR 8089 du CNRS,
Universit\'e de Cergy-Pontoise,
France, kitanine@ptm.u-cergy.fr\par
\hspace{1.5mm}
On leave of absence from Steklov Institute at
St. Petersburg, Russia},~~
{\bf J.~M.~Maillet}\footnote[2]{ Laboratoire de Physique, UMR 5672 du CNRS,
ENS Lyon,  France,
 maillet@ens-lyon.fr},~~
{\bf N.~A.~Slavnov}\footnote[3]{ Steklov Mathematical Institute,
Moscow, Russia, nslavnov@mi.ras.ru},~~
{\bf V.~Terras}\footnote[4]{LPMT, UMR 5825 du CNRS,
Montpellier, France, terras@lpm.univ-montp2.fr} \par

\end{large}

\vspace{80pt}

\centerline{\bf Abstract} \vspace{1cm}
\parbox{12cm}{\small%
We derive a new representation
for spin-spin correlation functions of the finite $XXZ$ spin-$1/2$ Heisenberg chain
in terms of a single multiple integral, that we call {\sl the master equation}.
Evaluation of this master equation gives rise on the one hand to the previously obtained
multiple integral formulas for the spin-spin
correlation functions and on the other hand to their expansion in
terms of the form factors of the local spin operators. Hence, it provides a
direct analytic link between
these two representations of  the correlation functions
and a complete re-summation of the corresponding series.
The  master equation method also allows one to obtain multiple integral
representations for dynamical correlation functions.
}
\end{center}

\newpage

\section{Introduction}
One of the central question in the theory of quantum integrable
models \cite{Bax82L,Fad84,GauL83,BogIK93L} is the exact
computation of their correlation functions. Apart from few cases,
like free fermions
\cite{Ons44,LieSM61,Mcc68,WuMTB76,MccTW77,SatMJ78} or conformal
field theories \cite{BelPZ84}, this problem is still far from its
complete solution. In particular, the computation of manageable
expressions for two point functions of local operators and their
asymptotic behavior at large distance is a central open problem.
If one considers the case $T=0$, such a problem reduces to the
calculation of the average value in the ground state
$|\omega\rangle$ of the product of two local operator
$\theta_1,\theta_2$:
\be{int-aver}
g_{12}=\langle\omega|\theta_1\theta_2|\omega\rangle.
\ee
There are basically two main strategies to evaluate such a function:

\vspace{2mm}

{\it (i)} to compute the action of local operators on the ground
state $\theta_1\theta_2|\omega\rangle = |\tilde\omega\rangle$ and
then
to calculate the resulting scalar product $g_{12}=\langle\omega|%
\tilde\omega\rangle$;

{\it (ii)} to insert a sum over a complete set of states
$|\omega_i\rangle$ (for instance, a complete set of eigenstates of
the Hamiltonian) between the local operators $\theta_1$ and
$\theta_2$ and to obtain the representation for the correlation
function as a sum over one-point matrix elements (form factor type
expansion \cite{KarW78,Smi92L,BabFKZ99})
\be{int-ff}
g_{12}=\sum_{i}\langle\omega|\theta_1|\omega_i\rangle\cdot\langle\omega_i|
\theta_2|\omega\rangle.
\ee
\vspace{2mm}

The aim of this paper is to give a direct, analytic relation
between the approaches {\it (i)} and {\it (ii)} in the case of the
$XXZ$ spin-$\frac12$ finite chain. For the sake of simplicity we
mainly consider the case of zero-temperature and equal-time
$\sigma^z$ correlation function. It will be clear however that our
method is more general. In particular, it  allows one to compute
dynamical correlation functions. The corresponding results will be
described in a sequel to the present paper.

We consider the periodical $XXZ$ spin-$\frac12$ Heisenberg chain in an external
magnetic field.
The Hamiltonian of this model is given by \cite{Hei28}
\be{IHamXXZ}
H=\sum_{m=1}^{M}\left(
\sigma^x_{m}\sigma^x_{m+1}+\sigma^y_{m}\sigma^y_{m+1}
+\Delta(\sigma^z_{m}\sigma^z_{m+1}-1)\right)-hS_z,
\ee
where
\be{ISz}
S_z=\frac{1}{2}\sum_{m=1}^{M}\sigma^z_{m},\qquad
[H,S_z]=0.
\ee
Here $\Delta$ is the anisotropy parameter, $h$ the external
classical magnetic field, $\sigma^{x,y,z}_{m}$ are the spin
operators (in the spin-$\frac12$ representation), associated with
each site of the chain $m$, and $M$ is even. The quantum space of
states is ${\cal H}={\otimes}_{m=1}^M {\cal H}_m$, where ${\cal
H}_m\sim \mathbb{C}^2$ is called the local quantum space at site
$m$. The operators $\sigma^{x,y,z}_{m}$ act as the corresponding
Pauli matrices in the space ${\cal H}_m$ and as the identity
operator elsewhere.

The method to compute eigenstates and energy levels (Bethe ansatz)
of the Hamiltonian \eq{IHamXXZ} was proposed by Bethe in 1931 in
\cite{Bet31} and developed later in \cite{Orb58,Wal59,YanY66a}.
The algebraic version of the Bethe ansatz was created in the
framework of the Quantum Inverse Scattering Method by L.D. Faddeev
and his school \cite{FadST79,TakF79,BogIK93L}. However the
knowledge of the correlation functions of the $XXZ$ chain has for
a long time been restricted to the free fermion point $\Delta = 0$.

In the case of the $XXZ$ chain the approach {\it (i)} leads to
multiple integral representations for the correlation functions.
In the thermodynamic limit, at zero temperature and for zero
magnetic field, such representations were obtained from the
$q$-vertex operator approach (also using corner transfer matrix
technique) in the massive regime $\Delta > 1$ in 1992
\cite{JimMMN92} and conjectured in 1996 \cite{JimM96} for the
massless regime $-1 < \Delta \le 1$ (see also \cite{JimML95}). A
proof of these results together with their extension to non-zero
magnetic field was given in 1999 \cite{KitMT99,KitMT00}
for both regimes using algebraic Bethe ansatz and the actual
resolution of the so-called quantum inverse scattering problem
\cite{KitMT99,MaiT99}. Using these results, the spontaneous
magnetization was obtained in \cite{IzeKMT99}. For the case of the
$XXX$ model ($\Delta=1$) this type of representation was later
studied in \cite{BosKS02,BosJMST04}. Integral representations for
spin-spin correlation functions were derived in
\cite{KitMST02a,KitMST02b}. Recently the generalization of this
result for finite temperature was given in \cite{GohKS04}.

In the framework of the approach {\it (ii)} integral
representations for the form factors of the $XXZ$ chain in the
thermodynamic limit were obtained in
\cite{JimMMN92,JimML95,Mik94,JimKMQ94,KojMQ95,Qua98}. Determinant
representations for the form factors of the finite chain were
computed in \cite{KitMT99,IzeKMT99}. An effective summation of the
form factor series in the case of free fermions was done in
\cite{BabB92,ColIKT93}.

To explain our method we will study  the two point correlation function of the third
components of spin $\langle\sigma_1^z\sigma_{m+1}^z\rangle$.
Following the papers \cite{IzeK85,ColIKT93,KitMST02a} we use 
for this purpose a special generating function
$\langle Q^\kappa_{1,m}\rangle$ (see \eq{GFdefQ}, \eq{exp-val}).
On the way to relate the previously obtained multiple integral representation
of the spin-spin correlation function
to its form factor type expansion we derive
what we call {\sl the master equation} (see \eq{Master-2})
for this generating function.
It is given as a single multiple integral of Cauchy type
over a certain contour $\Gamma$ in $\mathbb{C}^N$
\be{gen-res}
\langle Q^\kappa_{1,m}\rangle=
\oint\limits_{\Gamma}\,dz_1\dots dz_N\; F_\kappa(\{z\}).
\ee
The integrand $F_\kappa(\{z\})$ is a function of the $N$ variables
$z_1,\dots,z_N$. It is  mainly given in terms of the eigenvalues
$\tau_{\kappa}$ of the twisted transfer matrix (see \eq{TTM_def})
and of the functions ${\cal Y}_\kappa$ defining the corresponding
twisted Bethe equations (see \eq{TTM_Y-funct}, \eq{TTMBE_Y}). It is 
a periodical function of each argument $z_j$, vanishing at $z_j\to\pm\infty$.
Therefore there are two ways to evaluate the integral
\eq{gen-res}: either to compute the residues in the poles inside
$\Gamma$, or to compute the residues in the poles within
strips of the width $i\pi$ outside $\Gamma$.

The first way leads to a representation of the correlation function
$\langle\sigma_1^z\sigma_{m+1}^z\rangle$ in terms of the previously obtained
\cite{KitMST02a} $m$-multiple integrals.
Evaluation of the  integral \eq{gen-res} in terms of the poles
outside $\Gamma$ gives us the form factor type expansion of the correlation
function  (i.e. an  expansion in terms
of matrix elements of $\sigma^z$ between the ground state
and all excited states).

We would like to stress that the objects entering the above master
equation for the finite $XXZ$ chain are quite generic in the
context of quantum integrable models solvable by the algebraic
Bethe ansatz. Therefore we conjecture that similar formula holds
true in more general situations, in particular for the field
theory models. For these models the approach {\it (i)} usually
leads to short distance expansions for the correlation functions,
while the approach {\it (ii)} gives their long distance expansion.
The problem of putting into explicit (analytic) correspondence
these two regimes has been the subject of several works along the
last fifteen  years (see e.g.,
\cite{Zam91,Zam95,BabK03,BelBLPZ04}), remaining however up to now
an open problem. We hope that the method presented here could
ultimately shed some new light on these topics.

The paper is organized as follows. In the next section we give the
main notations and definitions of the XXZ model using the
framework of the algebraic Bethe ansatz. In Section 3 the twisted
transfer matrix ${\cal T}_\kappa$ is introduced. We describe the
properties of this operator and related objects, like  twisted
Bethe equations, their solutions and the scalar product formulas
for the eigenstates of ${\cal T}_\kappa$ with arbitrary vectors.
In Section 4 we derive the  master equation. The last two sections
are devoted to the above mentioned  two evaluations of the
multiple contour integral defining this  master equation. In
Section 5 we reproduce from it the multiple integral
representation for the $\sigma^z$ correlation function obtained in
\cite{KitMST02a}. This representation is given for the finite
chain and in the thermodynamic limit both for the massless and
massive regimes. In Section 6 we obtain the form factor type
expansion for the $\sigma^z$ correlation function. In the
Conclusion we give several comments on the key points of our
methods. We also discuss several perspectives of the obtained
result and announce the multiple integral representation for the
dynamical correlation function of the third components of spin,
which in fact is the subject of a forthcoming publication
\cite{KitMST04b}.


\section{Algebraic Bethe ansatz for the $XXZ$ chain}\label{B-f}

In the framework of the algebraic Bethe ansatz the Hamiltonian \eq{IHamXXZ}
can be obtained from the monodromy matrix $T(\lambda)$, that
in its turn is completely defined by the $R$-matrix. The
$R$-matrix of the $XXZ$ chain acts in the space
$\mathbb{C}^2\otimes \mathbb{C}^2$ and is equal to\footnote{%
Note that we use here  a different
normalization for the $R$-matrix compared to \cite{KitMST02a}.}
\be{ABAR}
R(\lambda)=\left(
\begin{array}{cccc}
\sinh(\lambda+\eta)&0&0&0\\
0&\sinh\lambda&\sinh\eta&0\\
0&\sinh\eta&\sinh\lambda&0\\
0&0&0&\sinh(\lambda+\eta)
\end{array}
\right),\qquad \cosh\eta=\Delta.
\ee
It is a solution of the Yang-Baxter equation. Identifying one of the two
vector spaces of the $R$-matrix with the quantum space ${\cal H}_m$, we
obtain the quantum $L$-operator at the site $m$
\be{ABL}
L_m(\lambda)=R_{0m}(\lambda-\eta/2).
\ee
Here $R_{0m}$ acts in $\mathbb{C}^2\otimes {\cal H}_m$. Then the monodromy
matrix $T(\lambda)$ is constructed as an ordered product of the $L$-operators
with respect to all the sites of the chain
\be{ABAT}
T(\lambda)=\left(
\begin{array}{cc}
A(\lambda)&B(\lambda)\\
C(\lambda)&D(\lambda)
\end{array}\right)=L_M(\lambda)
\dots L_2(\lambda) L_1(\lambda).
\ee
The operator $T(\lambda)$ acts in $V\otimes{\cal H}$, where $V\sim
\mathbb{C}^2$ is usually called auxiliary space of $T(\lambda)$.
The Hamiltonian \eq{IHamXXZ} at $h=0$ is related to $T(\lambda)$ by a
`trace identity'
\be{ABATI}
H=2\sinh\eta\left.\frac{d{\cal T}(\lambda)}{d\lambda}
{\cal T}^{-1}(\lambda)\right|_{\lambda=\frac{\eta}{2}}+const.
\ee
Here
\be{ABAtr}
{\cal T}(\lambda)=\tr T(\lambda)=A(\lambda)+D(\lambda).
\ee

Later on we shall consider the inhomogeneous $XXZ$
model, for which
\be{ABALinh}
L_m(\lambda)=L_m(\lambda,\xi_m)=R_{0m}(\lambda-\xi_m), \qquad
T(\lambda)=L_M(\lambda,\xi_M)
\dots L_2(\lambda,\xi_2)  L_1(\lambda,\xi_1),
\ee
where $\xi_m$ are arbitrary complex numbers attached to each lattice site that
are called inhomogeneity parameters. In the homogeneous limit $\xi_m=\eta/2$
and we come back to the original model.

The commutation relations between the entries of the monodromy matrix are
defined  by the Yang-Baxter quadratic relations,
\be{ABARTT}
R_{12}(\lambda_1-\lambda_2)T_1(\lambda_1)T_2(\lambda_2)=
T_2(\lambda_2)T_1(\lambda_1)R_{12}(\lambda_1-\lambda_2).
\ee
The equation \eq{ABARTT} holds in the space $V_1\otimes
V_2\otimes {\cal H}$ (where $V_j\sim \mathbb{C}^2$). The matrix
$T_j(\lambda)$ acts in a nontrivial way in the space $V_j\otimes {\cal H}$,
while the $R$-matrix is nontrivial in $V_1\otimes V_2$.

In the framework of the algebraic Bethe ansatz an arbitrary  quantum  state can be
obtained from the states generated  by
the action of the operators $B(\lambda)$ on the reference state $|0\rangle$
with all spins up,
\be{ABAES}
|\psi\rangle=\prod_{j=1}^{N}B(\lambda_j)|0\rangle,
\qquad N=0,1,\dots, M.
\ee
The eigenstates of the transfer matrix
${\cal T}(\mu)$ can be constructed in the form \eq{ABAES},
where the parameters $\lambda_j$ satisfy the system of
Bethe equations
\be{TTMBE}
a(\lambda_j)\prod_{k=1\atop{k\ne
j}}^{N} \sinh(\lambda_k-\lambda_j+\eta)=
d(\lambda_j)
\prod_{k=1\atop{k\ne
j}}^{N} \sinh(\lambda_k-\lambda_j-\eta),
\qquad j=1,\dots, N,
\ee
and $a(\lambda)$, $d(\lambda)$ are the eigenvalues of the operators
$A(\lambda)$ and $D(\lambda)$ on the reference state. In the normalization
\eq{ABAR}--\eq{ABL}, we have
\begin{align}\label{EV_DA}
a(\lambda) &=\prod_{a=1}^M\sinh(\lambda-\xi_a+\eta),\\
d(\lambda) &=\prod_{a=1}^M\sinh(\lambda-\xi_a).
\end{align}
{\sl Remark}. Generally speaking the system of equations \eq{TTMBE}
is neither a necessary nor a sufficient
condition for the vector \eq{ABAES} to be an eigenstate of the transfer
matrix. For example, the state \eq{ABAES} with all spins down ($N=M$)
is the eigenstate of ${\cal T}$ for generic $\{\lambda\}$.
On the other hand the system \eq{TTMBE} possesses solutions,
which do not correspond to any eigenstate of ${\cal T}$.
We discuss all these questions in more details in Section
\ref{TTM-SP} and Appendix \ref{Sol-TBE}.

The eigenvalue $\tau(\mu|\{\lambda\})$ of the operator
${\cal T}(\mu)$ corresponding to an eigenstate of the form \eq{ABAES} is
\be{ABAEV}
\tau(\mu|\{\lambda\})=
a(\mu)\prod_{k=1}^{N}\frac{\sinh(\lambda_k-\mu+\eta)}{\sinh(\lambda_k-\mu)}
+ d(\mu)
\prod_{k=1}^{N}\frac{\sinh(\mu-\lambda_k+\eta)}{\sinh(\mu-\lambda_k)}~,
\ee
The dual states can be constructed
similarly to \eq{ABAES} via the operators $C(\lambda)$
\be{ABADES}
\langle\psi|=\langle 0|\prod_{j=1}^{N}C(\lambda_j),
\qquad N=0,1,\dots, M.
\ee
Here  $\langle 0|=|0\rangle^{+}$ and the dual
eigenstates of ${\cal T}(\mu)$ are given in the form
\eq{ABADES}, where the parameters $\lambda_j$
satisfy the same system of equations \eq{TTMBE}. Generically
$\langle\psi|\ne|\psi\rangle^\dagger$ due
to the involution $C(\lambda)=\pm B^\dagger(\bar\lambda)$.
If, however, the state \eq{ABAES} is the ground state
of the Hamiltonian \eq{IHamXXZ},
then $\langle\psi|=c^N|\psi\rangle^\dagger$ with $c=1$
for $\Delta>1$ and  $c=-1$ for $-1<\Delta\le1$ respectively.

In the end of this section we give the explicit representations
of the local spin operators in terms of the entries of the monodromy 
matrix. Such a representation is given by the solution of the quantum 
inverse scattering problem \cite{KitMT99,MaiT99}
\be{FCtab}
\sigma^\alpha_j=\prod_{k=1}^{j-1}{\cal T}(\xi_k)\cdot
\tr\bigl(T(\xi_j)\sigma^\alpha\bigr)
\cdot\prod_{k=1}^{j}{\cal T}^{-1}(\xi_k).
\ee
Here $\sigma^\alpha$ in the r.h.s. acts in the auxiliary space of
$T(\lambda)$, while $\sigma^\alpha_j$ in the l.h.s. acts in the
local quantum space ${\cal H}_j$.
The formulas of the quantum inverse scattering problem permit us
to embed the problem of calculation of the correlation functions
of the local spin operators into the algebraic Bethe ansatz.


\section{Twisted transfer matrix and scalar products}\label{TTM-SP}

In this section we introduce new objects and notations, which
will be used through all the paper. Starting from this section
we consider only the subspace ${\cal H}^{(M/2-N)}$ of the quantum space
${\cal H}$ with fixed (but arbitrary) number of spins down $N$.

First, we introduce the `twisted transfer matrix' ${\cal T}_\kappa(\mu)$
\be{TTM_def}
{\cal T}_\kappa(\mu)=A(\mu)+\kappa D(\mu),
\ee
where $\kappa$ is a complex parameter. We denote also as $|\psi_\kappa
(\{\lambda\})\rangle$ and $\langle\psi_\kappa(\{\lambda\})|$ the eigenstates
(respectively the dual eigenstates) of the operator ${\cal T}_\kappa(\mu)$
in the subspace ${\cal H}^{(M/2-N)}$. The corresponding
eigenvalue is denoted as $\tau_\kappa(\mu|\{\lambda\})$. The same notations
without a subscript $\kappa$ correspond to the case $\kappa=1$ considered in
the previous section.

The eigenstates  of ${\cal T}_\kappa(\mu)$
(and their dual states) have the form
\be{ABA-TES}
|\psi_\kappa(\{\lambda\})\rangle
=\prod_{j=1}^{N}B(\lambda_j)|0\rangle,\qquad
\langle\psi_\kappa(\{\lambda\})|=\langle 0|\prod_{j=1}^{N}C(\lambda_j),
\ee
with an eigenvalue
\be{T-ev}
\tau_\kappa(\mu|\{\lambda\})=a(\mu)\prod_{k=1}^{N}
\frac{\sinh(\lambda_k-\mu+\eta)}{\sinh(\lambda_k-\mu)}
+\kappa\, d(\mu)\prod_{k=1}^{N}\frac{\sinh(\mu-\lambda_k+\eta)}
{\sinh(\mu-\lambda_k)}~,
\ee
where the parameters $\{\lambda\}$ satisfy the system of twisted  Bethe
equations
\be{TBE-l-1}
a(\lambda_j)\prod_{k=1\atop{k\ne
j}}^{N} \sinh(\lambda_k-\lambda_j+\eta)=
\kappa\,d(\lambda_j)
\prod_{k=1\atop{k\ne
j}}^{N} \sinh(\lambda_k-\lambda_j-\eta),
\qquad j=1,\dots, N.
\ee
It is also convenient to introduce the function
\begin{align}
{\cal Y}_\kappa(\mu|\{\lambda\}) &=
\prod_{k=1}^{N}\sinh(\lambda_k-\mu)\cdot
\tau_\kappa(\mu|\{\lambda\})\nonumber
\\
    &=a(\mu) \prod_{k=1}^{N}\sinh(\lambda_k-\mu+\eta) +\kappa\, d(\mu)
\prod_{k=1}^{N}\sinh(\lambda_k-\mu-\eta)~.\label{TTM_Y-funct}
\end{align}
In terms of this function the system \eq{TBE-l-1}
reads
\be{TTMBE_Y}
{\cal Y}_\kappa(\lambda_j|\{\lambda\})=0, \qquad j=1,\dots,N.
\ee

Just like in the case $\kappa=1$, for generic $\kappa$ not all  the
solutions of the system \eq{TTMBE_Y}  correspond to eigenvectors
of ${\cal T}_\kappa(\mu)$. A brief sketch of the properties of the
solutions of \eq{TTMBE_Y} is given in Appendix \ref{Sol-TBE}
(for details we
refer the reader to the original paper \cite{TarV96}). Here we merely
recall some definitions and one of the main results of
\cite{TarV96}.

\begin{Def}
A solution $\{\lambda\}$ of the system \eq{TTMBE_Y} is called admissible, if
\be{admiss}
d(\lambda_j)\prod_{k=1\atop{k\ne j}}^N\sinh(\lambda_j-\lambda_k+\eta)\ne0,\qquad
j=1,\dots,N,
\ee
and unadmissible otherwise. A solution is called off-diagonal
if the parameters $\{\lambda\}$ are pair-wise distinct and
diagonal otherwise. A solution is called  degenerated, if the
Jacobian of the system \eq{TTMBE_Y} vanishes, and  non-degenerated
otherwise. A solution is called trivial, if the  corresponding
state $|\psi_\kappa(\{\lambda\})\rangle$ is the zero vector.
\end{Def}

One of the main result of \cite{TarV96} is that, for generic
$\kappa$ and $\{\xi\}$, the eigenstates of ${\cal T}_\kappa(\mu)$
corresponding to the admissible off-diagonal solutions of \eq{TTMBE_Y}
form a basis in the space ${\cal H}^{(M/2-N)}$.
Generically this basis is not normalized and non-orthogonal,
although it is orthogonal to the dual basis.
We would like to mention that for some specific choice of
$\kappa$ and $\{\xi\}$ certain unadmissible solutions may also contribute
to the basis of the eigenstates  in ${\cal H}^{(M/2-N)}$.

Consider now the scalar products of eigenstates \eq{ABA-TES} and arbitrary
states of the form \eq{ABAES}. The explicit results for such scalar products
at $\kappa=1$ were obtained in \cite{Sla89,KitMT99}. Applying the methods
used in these papers one can easily prove
\begin{prop}
Let $\{\lambda\}$ satisfy the system \eq{TTMBE_Y},
$\{\mu\}$ be generic complex numbers. Then
\begin{multline}\label{FCdet}
\langle 0|\prod_{j=1}^{N}C(\mu_j)|\psi_\kappa(\{\lambda\})\rangle
= \langle\psi_\kappa(\{\lambda\})|\prod_{j=1}^{N}B(\mu_j)|0\rangle
\\
=\prod_{a=1}^{N} d(\lambda_a)\cdot
    {\cal X}_N^{-1}(\{\mu\},\{\lambda\})\cdot
   \det_N \left(\frac\partial{\partial\lambda_j}
   \tau_\kappa(\mu_k|\{\lambda\})\right).
\end{multline}
Here ${\cal X}_N(\{\mu\},\{\lambda\})$ is the Cauchy determinant composed
of the parameters $\{\lambda\}$ and $\{\mu\}$
\be{TTM_Cauchy}
{\cal X}_N(\{\mu\},\{\lambda\})={\det}_N\left(
\frac1{\sinh(\mu_k-\lambda_j)}\right)=
\frac{\prod\limits_{a>b}^N\sinh(\lambda_a-\lambda_b)
\sinh(\mu_b-\mu_a)}{\prod\limits_{a,b=1}^N\sinh(\mu_b-\lambda_a)}.
\ee
\end{prop}

To make these formulas more explicit we introduce for arbitrary positive
integers $n$ and $n'$ ($n\le n'$) and
arbitrary sets of variables $\lambda_1,\dots,\lambda_n$, $\mu_1,%
\dots,\mu_n$ and $\nu_1,\dots,\nu_{n'}$,
such that $\{\lambda\} \subset \{\nu\}$,  the following $n\times n$ matrix
$\Omega_\kappa(\{\lambda\},\{\mu\}|\{\nu\})$ 
\begin{multline} \label{matH}
  (\Omega_\kappa)_{jk}(\{\lambda\},\{\mu\}|\{\nu\})=
  a(\mu_k)\,t(\lambda_j,\mu_k)\,\prod_{a=1}^{n'} \sinh(\nu_a-\mu_k+\eta)\\
   -\kappa\, d(\mu_k)\,t(\mu_k,\lambda_j)\,\prod_{a=1}^{n'} \sinh(\nu_a-\mu_k-\eta),
\end{multline}
with
\be{def-t}
t(\lambda,\mu)=\frac{\sinh\eta}{\sinh(\lambda-\mu)\sinh(\lambda-\mu+\eta)}.
\ee
Then the equation \eq{FCdet} reads
\begin{multline}\label{m-l-H}
\langle 0|\prod_{j=1}^{N}C(\mu_j)|\psi_\kappa(\{\lambda\})\rangle
=\langle\psi_\kappa(\{\lambda\})|\prod_{j=1}^{N}B(\mu_j)|0\rangle
\\
=\frac{\prod_{a=1}^{N} d(\lambda_a)}
{\prod\limits_{a>b}^N\sinh(\lambda_a-\lambda_b)\sinh(\mu_b-\mu_a)}
\cdot \det_N \Omega_\kappa(\{\lambda\},\{\mu\}|\{\lambda\}).
\end{multline}
Taking the limit $\{\mu\}\to\{\lambda\}$ in \eq{m-l-H} we obtain the
'square of the norm' of the eigenstate (recall that generically
$\langle\psi_\kappa(\{\lambda\})|\ne|\psi_\kappa(\{\lambda\})
\rangle^\dagger$) \cite{GauMW81,Kor82}
\be{norm}
\langle \psi_\kappa(\{\lambda\})|\psi_\kappa(\{\lambda\})\rangle
= \frac{\prod_{a=1}^{N} d(\lambda_a)}
{\prod\limits_{a,b=1\atop{a\ne b}}^N\sinh(\lambda_a-\lambda_b)}
\cdot \det_N \Omega_\kappa(\{\lambda\},\{\lambda\}|\{\lambda\}).
\ee
This equation can also be written in terms of a Jacobian using
\be{H-Jac}
\det_N\Omega_\kappa(\{\lambda\},\{\lambda\}|\{\lambda\})=
\det_N\left(-\frac\partial{\partial\lambda_k}
{\cal Y}_\kappa(\lambda_j|\{\lambda\})\right),
\ee
where $\{\lambda\}$  satisfy the system \eq{TTMBE_Y}.


\section{Master equation for $\sigma^z$ correlation function}\label{D-ME}

In this paper we consider the generating functional for the correlation
function of the third components of spins
$\langle\sigma_1^z\sigma_{m+1}^z\rangle$. Following the papers \cite{IzeK85,%
ColIKT93} we define for any complex number
$\kappa$ the operator $Q^\kappa_{1,m}$ as\footnote[1]{%
Setting $\kappa=e^\beta$ one has $Q^\kappa_{1,m}=\exp\left(
\frac\beta2\sum_{n=1}^{m}(1-\sigma_n^z)\right)$ and, hence, this is exactly
the operator considered in \cite{KitMST02a}.}
\be{GFdefQ}
Q^\kappa_{1,m}=\prod_{n=1}^m\left(\frac{1+\kappa}2+\frac{1-\kappa}2\cdot
\sigma_n^z\right).
\ee
The generating functional is equal to
the expectation value
\be{exp-val}
\langle Q^\kappa_{1,m}\rangle
=\frac{\langle\psi(\{\lambda\})|Q^\kappa_{1,m}
|\psi(\{\lambda\})\rangle}{\langle\psi(\{\lambda\})|\psi(\{\lambda\})\rangle},
\ee
where $|\psi(\{\lambda\})\rangle$ is an eigenstate of ${\cal T}(\mu)$
in the subspace ${\cal H}^{(M/2-N)}$.
Taking the second `lattice derivative' of
$\langle Q^\kappa_{1,m}\rangle$ and the
second derivative with respect to $\kappa$ at $\kappa=1$
we extract the two-point correlation function of the
third components of local spins:
\be{GFlatder}
\frac12\langle(1-\sigma_1^z)(1-\sigma_{m+1}^z)\rangle=
\left.\frac{\partial^2}{\partial\kappa^2}
\langle \left(Q_{1,m+1}^\kappa-Q_{1,m}^\kappa-
Q_{2,m+1}^\kappa+ Q_{2,m}^\kappa\right)\rangle
\right|_{\kappa=1}.
\ee

The main result of this paper is an integral representation for
the generating functional \eq{exp-val}.

\begin{thm}\label{M-thm} Let the inhomogeneities $\{\xi\}$ be generic
and  the set $\{\lambda\}$ be an
admissible off-diagonal solution of the system \eq{TTMBE}.
Then there exists $\kappa_0>0$ such, that for $|\kappa|<\kappa_0$
the expectation value of the operator
$Q^\kappa_{1,m}$ in the inhomogeneous finite $XXZ$
chain is given by the multiple contour integral
\begin{multline}\label{Master-2}
\langle Q^\kappa_{1,m}\rangle=
\frac{1}{N!}
 \oint\limits_{\Gamma\{\xi\} \cup \Gamma\{\lambda\}}
 \prod_{j=1}^{N}\frac{dz_j}{2\pi i} \cdot
\prod_{a,b=1}^{N}\sinh^2(\lambda_a-z_b)
\cdot\prod_{a=1}^m\frac{\tau_\kappa(\xi_a|\{z\})}
{\tau(\xi_a|\{\lambda\})}
\\
\times\frac{\det_N
\left(\frac{\partial\tau_\kappa(\lambda_j|\{z\})}{\partial z_k}\right)
\cdot\det_N
\left(\frac{\partial\tau(z_k|\{\lambda\})}{\partial \lambda_j}\right)}
{\prod\limits_{a=1}^{N}{\cal Y}_\kappa (z_a| \{z\} )
\cdot\det_N
\left(\frac{\partial{\cal Y}(\lambda_k|\{\lambda\})}
{\partial\lambda_j}\right)}.
\end{multline}
The integration contour is such that the only singularities of
the integrand \eq{Master-2} within
$\Gamma\{\xi\}\cup \Gamma\{\lambda\}$ which contribute to the
integral  are the points $\{\xi\}$ and $\{\lambda\}$.
\end{thm}

We call \eq{Master-2} {\sl  the master equation}.

{\sl Remark 1}. The master equation \eq{Master-2}
gives the expectation value $\langle Q^\kappa_{1,m}\rangle$
with respect to an arbitrary eigenstate $|\psi(\{\lambda\})\rangle$ of
${\cal T}$. In particular one can chose $\{\lambda\}$ such that
in the homogeneous limit $\xi_a\to\eta/2$ the corresponding eigenstate
$|\psi(\{\lambda\})\rangle$ goes to the ground state of the
$XXZ$ Hamiltonian.

{\sl Remark 2}. The master equation \eq{Master-2} is
an integral representation of the expectation value
$\langle Q^\kappa_{1,m}\rangle$, which holds true
at least for $|\kappa|$ small enough. On the other hand
this expectation value is evidently a polynomial in $\kappa$
of degree $m$. Therefore the representation \eq{Master-2}
can be easily continued in $\kappa$ from any vicinity of the origin to the whole
complex plane. This does not mean, however, that one can set $\kappa$ to
be an arbitrary specific value directly in the integrand of \eq{Master-2}.

{\sl Proof of Theorem \ref{M-thm}}. In order to compute the expectation
value of $Q^\kappa_{1,m}$, we express this operator in terms of the twisted
transfer matrix. Due to \eq{FCtab} one has
\be{GFexpbQ}
Q^\kappa_{1,m}=\prod_{a=1}^{m}{\cal T}_\kappa(\xi_a)
\prod_{b=1}^{m}{\cal T}^{-1}(\xi_b).
\ee
Then we can use the explicit formula for the
multiple action of $\prod_{a=1}^m{\cal T}_\kappa(x_a)$ for an
arbitrary set of complex numbers $\{x\}$
on an arbitrary state $\langle0|\prod_{j=1}^NC(\lambda_j)$,
\cite{KitMST02a}
 \begin{multline}\label{APcompl-act}
 \langle0|\prod_{j=1}^NC(\lambda_j)\prod_{a=1}^m{\cal T}_\kappa(x_a)
    \\
 =\sum_{n=0}^{\min{(m,N)}}
 \sum_{\{\lambda\}=\{\lambda_{\alpha_+}\}\cup\{\lambda_{\alpha_-}\}
 \atop{\{x\}=\{x_{\gamma_+}\}\cup\{x_{\gamma_-}\}
 \atop{|\alpha_+|=|\gamma_+|=n}}}
 R_n^\kappa(\{x_{\gamma_+}\},\{x_{\gamma_-}\},
\{\lambda_{\alpha_+}\},\{\lambda_{\alpha_-}\})\,
 \langle 0 |\prod_{a\in\gamma_+} C(x_a)
 \prod_{b\in\alpha_-}C(\lambda_b),
\end{multline}
where the coefficient $R_n^\kappa(\{x_{\gamma_+}\},\{x_{\gamma_-}\},
\{\lambda_{\alpha_+}\},\{\lambda_{\alpha_-}\})$ is given by
 \begin{multline}\label{APRn}
 R_n^\kappa =
\biggl\{\prod_{a>b \atop a,b\in\alpha_+}\sinh(\lambda_b-\lambda_a)
\prod_{a<b \atop a,b\in\gamma_+}\sinh(x_b-x_a)
\prod_{a\in\alpha_+}\prod_{b\in\alpha_-}\sinh(\lambda_b-\lambda_a)
\biggr\}^{-1}\\
\times \prod_{a\in\gamma_-}\tau_\kappa(x_a|\{x_{\gamma_+}\}\cup
\{\lambda_{\alpha_-}\})  \cdot
\det_n \Omega_\kappa(\{x_{\gamma_+}\},\{\lambda_{\alpha_+}\}|\{x_{\gamma_+}\}\cup
 \{\lambda_{\alpha_-}\}).
 \end{multline}
Here
$\Omega_\kappa(\{x_{\gamma_+}\},\{\lambda_{\alpha_+}\}|\{x_{\gamma_+}\}\cup
 \{\lambda_{\alpha_-}\})$ is  given by expressions \eq{matH}
in which the sets $\{\lambda\}$, $\{\mu\}$, $\{\nu\}$ have to be
replaced by $\{x_{\gamma_+}\}$, $\{\lambda_{\alpha_+}\}$ and
$\{x_{\gamma_+}\}\cup \{\lambda_{\alpha_-}\}$ respectively.

The summation in \eq{APcompl-act} is taken with respect to
partitions of the sets $\{\lambda\}$ and $\{x\}$ into disjoint
subsets
$\{\lambda\}=\{\lambda_{\alpha_+}\}\cup\{\lambda_{\alpha_-}\}$ and
$\{x\}=\{x_{\gamma_+}\}\cup\{x_{\gamma_-}\}$, such that the number of
elements in the subsets $\{\lambda_{\alpha_+}\}$ and
$\{x_{\gamma_+}\}$ coincides and is equal to $n$.

In the paper \cite{KitMST02a} we directly applied the equation \eq{APcompl-act}
for the computation of $\langle Q^\kappa_{1,m}\rangle$, specifying
$x_j=\xi_j$ for $j=1,\dots,m$. The peculiarity of the inhomogeneity parameters is
that $d(\xi_j)=0$, what simplifies the formulas. In
the present paper, however, we are going to keep the parameters $\{x\}$
arbitrary, i.e. we shall consider the expectation value of the operator
\be{GFexp_tQ}
{\cal Q}_m(\kappa,\{x\})=\prod_{a=1}^{m}{\cal T}_\kappa(x_a)\cdot
\prod_{b=1}^{m}{\cal T}^{-1}(x_b),
\ee
where the  parameters $x_1,\dots,x_m$ are generic complex numbers.

The normalized expectation value
of ${\cal Q}_m(\kappa,\{x\})$ on the finite lattice with respect to an
arbitrary eigenstate of the transfer matrix
$|\psi(\{\lambda\})\rangle$ has the form
\be{GFexpbQev}
   \langle {\cal Q}_m(\kappa,\{x\}) \rangle
  =\frac{\langle\psi(\{\lambda\})|\prod\limits_{a=1}^{m}
{\cal T}_\kappa(x_a)\cdot
\prod\limits_{b=1}^{m}{\cal T}^{-1}(x_b)|\psi(\{\lambda\})\rangle}
   {\langle\psi(\{\lambda\})|\psi(\{\lambda\})\rangle}.
\ee
The action to the right of the product of ${\cal T}^{-1}(x_b)$
produces merely a numerical factor:
\be{GFactright}
\prod\limits_{b=1}^{m}{\cal T}^{-1}(x_b)|\psi(\{\lambda\})\rangle=
\prod\limits_{b=1}^{m}\tau^{-1}(x_b|\{\lambda\})\cdot
|\psi(\{\lambda\})\rangle.
\ee
Acting to the left with the product of ${\cal T}_\kappa(x_a)$
by means of \eq{APcompl-act} with $m<N$, and using the expression \eq{m-l-H} of the
scalar product at $\kappa=1$, we obtain
\begin{equation}\label{eq-del}
  \langle {\cal Q}_m(\kappa,\{x\}) \rangle
    = \sum_{n=0}^{m}
      \sum_{\{\lambda\}=\{\lambda_{\alpha_+}\}\cup\{\lambda_{\alpha_-}\}
 \atop{\{x\}=\{x_{\gamma_+}\}\cup\{x_{\gamma_-}\}
 \atop{|\alpha_+|=|\gamma_+|=n}}}
 \Delta_n^\kappa(\{x_{\gamma_+}\},\{x_{\gamma_-}\},\{\lambda_{\alpha_+}\}
,\{\lambda_{\alpha_-}\}),
\end{equation}
with
\begin{multline}
   \Delta_n^\kappa =
     \prod_{b=1}^m \Bigl\{\tau(x_b|\{\lambda\})
\prod\limits_{a\in\alpha_-} \sinh(\lambda_a-x_b)
    \prod\limits_{a\in\gamma_+\atop a\ne b}\sinh(x_a-x_b)\Bigr\}^{-1}
    \prod\limits_{a\in\gamma_-}
              {\cal Y}_\kappa(x_a|\{x_{\gamma_+}\}\cup\{\lambda_{\alpha_-}\})
\\
    \times
    \det_n \Omega_\kappa(\{ x_{\gamma_+} \},\{ \lambda_{\alpha_+} \}
                           | \{ x_{\gamma_+} \} \cup \{\lambda_{\alpha_-}\} )
       \cdot
    \frac{ \det_N \Omega (\{\lambda\},
                \{x_{\gamma_+}\} \cup \{\lambda_{\alpha_-}\}|\{\lambda\})}
{\det_N\Omega(\{\lambda\},\{\lambda_{\alpha_+}\} \cup \{\lambda_{\alpha_-}\}
| \{\lambda\})}.
\end{multline}
Here ${\cal Y}$ and $\Omega$ without subscript $\kappa$
are equal respectively to ${\cal Y}_\kappa$ and $\Omega_\kappa$ at $\kappa=1$.
The elements in the sets
$\{x_{\gamma_+}\} \cup \{\lambda_{\alpha_-}\}$ and
$\{\lambda_{\alpha_+}\} \cup \{\lambda_{\alpha_-}\}$ are ordered
accordingly.

Like in \cite{KitMST02a}, we can now perform a re-summation over
the partitions of the set $\{x\}$ by introducing contour integrals
over auxiliary variables $z_1,\dots,z_n$:
\begin{multline}\label{resum-x}
 \langle {\cal Q}_m(\kappa,\{x\}) \rangle =
 \sum_{n=0}^{m}
 \sum_{\{\lambda\}=\{\lambda_{\alpha_+}\}\cup\{\lambda_{\alpha_-}\}
 \atop{|\alpha_+|=n}}
 \prod_{a=1}^{m}\prod\limits_{b\in\alpha_-}\sinh^{-1}(\lambda_b-x_a)
 \cdot\frac{1}{n!}\oint\limits_{\Gamma\{x\}}
 \prod_{j=1}^{n}\frac{dz_j}{2\pi i}\\
 \times
 \prod_{a=1}^{n}\prod_{b=1}^m \frac{1}{\sinh(z_a-x_b)}
 \cdot
 \prod_{a=1}^{m}\frac{ {\cal Y}_\kappa (x_a| \{z\} \cup \{\lambda_{\alpha_-}\} ) }
                     { \tau (x_a| \{\lambda\} ) } \cdot
 \prod_{a=1}^{n} \frac{1}
                      { {\cal Y}_\kappa (z_a| \{z\} \cup \{\lambda_{\alpha_-}\} )}
         \\
 \times \det_n \Omega_\kappa ( \{z\} , \{\lambda_{\alpha_+}\}
                           | \{z\} \cup \{\lambda_{\alpha_-}\} )
    \cdot
    \frac{ \det_N \Omega (\{\lambda\} , \{z\} \cup \{\lambda_{\alpha_-}\}
                                                                  |\{\lambda\} )}
{\det_N\Omega (\{\lambda\} ,\{\lambda_{\alpha_+}\} \cup \{\lambda_{\alpha_-}\}
| \{\lambda\})},
\end{multline}
where the closed contour $\Gamma\{x\}$  surrounds the points\footnote[1]{%
More precisely, $\Gamma\{x\}$ is the boundary of a set of
polydisks ${\cal D}_a(r)$  in $\mathbb{C}^n$. Namely,
$\Gamma\{x\}={\cup}_{a=1}^m\bar{\cal D}_a(r)$, where $\bar{\cal
D}_a(r)=\{z\in\mathbb{C}^n: |z_k-x_a|=r,\quad k=1,\dots,n\}$. The
integration contour $\Gamma\{\xi\}\cup\Gamma\{\lambda\}$ in
\eq{Master-2} should be
understood in a similar manner in $\mathbb{C}^N$. }
$x_1,\dots,x_m$ and do not contain any other pole of the integrand, i.e. zeros in
$\mathbb{C}^n$ of the functions ${\cal Y}_\kappa(z_a|\{z\}
\cup \{\lambda_{\alpha_-}\})$. Since at this stage of the computations
$x_1,\dots,x_m$ are generic complex numbers, we can always choose them 
separated from the zeros of ${\cal Y}_\kappa(z_a|\{z\}\cup \{\lambda_{\alpha_-}\})$
for any subset $\{\lambda_{\alpha_-}\}$.

Thus, the sum over partitions of the set $\{x\}$ in \eq{eq-del} is
replaced with a contour integral. In paper \cite{KitMST02a} we
replaced in the thermodynamical limit
the sum over partitions of the set $\{\lambda\}$ with the integrals
over the support of the spectral density of the ground state. Here we treat
this sum in a different way. Namely, we perform a second re-summation over the
partitions of the set $\{\lambda\}$
in \eq{resum-x} also in terms of a contour integral.
Indeed, one can easily see that
\begin{multline}\label{resH}
   \Res_{\{z_{n+1},\dots,z_N\}=\{\lambda_{\alpha_-}\}}
     \Bigl[ \det_N \Omega_\kappa ( \{z_1,\dots,z_N\} ,
 \{\lambda_{\alpha_+}\} \cup \{\lambda_{\alpha_-}\}|\{z_1,\dots,z_N\})\Bigr]\\
   = \prod_{a\in\alpha_-}
    {\cal Y}_\kappa (\lambda_a| \{z_1,\dots,z_n\} \cup \{\lambda_{\alpha_-}\} )
     \cdot
    \det_n \Omega_\kappa ( \{z_1,\dots,z_n\} , \{\lambda_{\alpha_+}\}
                           |\{z_1,\dots,z_n\} \cup \{\lambda_{\alpha_-}\}).
\end{multline}
This enables us to express the generating functional
$\langle {\cal Q}_m(\kappa,\{x\}) \rangle$ for the finite
$XXZ$ chain as a
single multiple integral:
\begin{multline}\label{master1}
 \langle {\cal Q}_m(\kappa,\{x\}) \rangle =
 \frac{1}{N!}
 \oint\limits_{\Gamma\{x\} \cup \Gamma\{\lambda\}}
\prod_{j=1}^N\frac{dz_j}{2\pi i}\cdot
 \prod_{a=1}^{m}\frac{ \tau_\kappa (x_a| \{z\} ) }
                     { \tau (x_a| \{\lambda\} ) }\cdot
\prod_{a=1}^{N} \frac{1}
                      { {\cal Y}_\kappa (z_a| \{z\} )}
\\
 \times
 \det_N \Omega_\kappa ( \{z\} , \{\lambda\} | \{z\}  )
    \cdot
    \frac{ \det_N \Omega (\{\lambda\} , \{z\}  |\{\lambda\} )}
         { \det_N \Omega (\{\lambda\} , \{\lambda\} | \{\lambda\})},
\end{multline}
where the closed contour $\Gamma\{x\} \cup \Gamma\{\lambda\}$
surrounds the points $x_1,\dots,x_m$  and $\lambda_1,\dots,\lambda_N$ (spectral
parameters corresponding to given eigenstate of ${\cal T}(\mu)$) and no other
poles of the integrand. 

Finally, in order to reproduce the expectation value $\langle Q^\kappa_{1,m},
\rangle$ we should set the generic parameters $\{x\}$
in \eq{master1} to be equal to the inhomogeneities $\xi_1,\dots,\xi_m$.
This definitely can be done if the system
\be{TBE-z}
{\cal Y}_\kappa (z_j|\{z\})=0,\qquad j=1,\dots,N
\ee
has no solution inside the
integration contour $\Gamma\{\xi\}\cup\Gamma\{\lambda\}$. Thus,
we need to analyze the properties of the solutions of the system
of the twisted Bethe equations. Hereby,
since $\langle {\cal Q}_m(\kappa,\{x\}) \rangle$ is a polynomial in
$\kappa$, it is sufficient to determine this polynomial in an open ball
around some specific value, for example
for $|\kappa|$ small enough.

The detail analysis of the system \eq{TBE-z} was done in the
paper \cite{TarV96}. We give some of the basic statements of this
paper in Appendix \ref{Sol-TBE}. Here we present only the 
results necessary for
writing and evaluating the integral \eq{master1}.

\begin{lemma}\label{main-lem-0}
Let the inhomogeneities $\xi_1,\dots,\xi_m$ be generic, 
$\lambda_1,\dots, \lambda_N$ be an admissible off-diagonal
solution of \eq{TTMBE} and $|\kappa|$ be small enough. Then all
admissible off-diagonal solutions of the system \eq{TBE-z} are
separated from the points $\{\xi\}$ and $\{\lambda\}$.
\end{lemma}

\begin{cor}\label{corol}
There exists a contour $\Gamma\{\xi\}\cup\Gamma\{\lambda\}$ such
that the points $\{\xi\}$ and $\{\lambda\}$ are inside this
contour, while all admissible off-diagonal solutions of the system
\eq{TBE-z} are outside $\Gamma\{\xi\}\cup\Gamma\{\lambda\}$.
\end{cor}

\begin{lemma}\label{main-lem}
Let the contour $\Gamma\{\xi\}\cup\Gamma\{\lambda\}$ satisfy
the conditions of the corollary \ref{corol}, and $x_k\to\xi_k$. Then:

1) the only poles
inside the contour $\Gamma\{\xi\}\cup\Gamma\{\lambda\}$ which
provide non-vanishing contribution to the integral \eq{master1}
are in $\{\xi\}\cup\{\lambda\}$;

2) the only poles outside the contour
$\Gamma\{\xi\}\cup\Gamma\{\lambda\}$ which provide non-vanishing
contribution to the integral \eq{master1} are the admissible
off-diagonal solutions of the system \eq{TBE-z}.
\end{lemma}

The proof of these lemmas is given in the Appendix \ref{Sol-TBE} using the results
of \cite{TarV96}.

Lemma \ref{main-lem} guarantees that for $x_k=\xi_k$
there are no any other poles of the integrand \eq{master1} inside the contour
$\Gamma\{\xi\}\cup\Gamma\{\lambda\}$ and contributing to
the integral except $z_j=\xi_k$ and
$z_j=\lambda_k$. Thus,  setting $x_k=\xi_k$ and using $d(\xi_k)=0$
we arrive at
\begin{multline}\label{Master-1}
\langle Q^\kappa_{1,m}\rangle=
\frac{1}{N!}
 \oint\limits_{\Gamma\{\xi\} \cup \Gamma\{\lambda\}}
 \prod_{j=1}^{N}\frac{dz_j}{2\pi i} \cdot
 \prod_{a=1}^{N}\prod_{b=1}^m
\left(\frac{\sinh(\lambda_a-\xi_b)}{\sinh(z_a-\xi_b)}
 \cdot\frac{\sinh(z_a-\xi_b+\eta)}{\sinh(\lambda_a-\xi_b+\eta)}\right)
\\
 \times\frac{\det_N \Omega_\kappa ( \{z\} , \{\lambda\} | \{z\})
\cdot \det_N \Omega (\{\lambda\} , \{z\}  |\{\lambda\} )}
{\prod\limits_{a=1}^{N}{\cal Y}_\kappa (z_a| \{z\} )\cdot
\det_N \Omega (\{\lambda\} , \{\lambda\} | \{\lambda\})},
\end{multline}
It remains to express the determinants of the matrices
$\Omega_\kappa$ and $\Omega$ in terms of Jacobians using
\eq{FCdet}--\eq{H-Jac}, and we obtain the master equation
\eq{Master-2}.$\Box$


\section{Multiple integral representation}\label{S-D-E}

Lemma \ref{main-lem} permits us to evaluate the integral
\eq{Master-2} in two different ways: either to compute the residues
of the integrand inside the contour
$\Gamma\{\xi\}\cup\Gamma\{\lambda\}$, or to compute the residues
outside this contour. In this section we consider the first way,
which immediately leads us to
the representation for $\langle Q^\kappa_{1,m}\rangle$ obtained
in \cite{KitMST02a}.

For this purpose it is more convenient to use \eq{Master-1}, which in fact
is equivalent to \eq{Master-2}. This multiple integral
can be presented as
\be{dec-integr}
\oint\limits_{\Gamma\{\xi\} \cup \Gamma\{\lambda\}}\prod_{j=1}^N
\,dz_j=\sum_{n=0}^N C_N^n
\oint\limits_{\Gamma\{\xi\}}\prod_{j=1}^n\,dz_j
\oint\limits_{\Gamma\{\lambda\}}\prod_{j=1}^{N-n}\,dz_j.
\ee
Hereby, since the number of the poles inside ${\Gamma\{\xi\}}$ is
$m$ and the integrand vanishes as soon as $z_j=z_k$, the sum in
\eq{dec-integr} is actually restricted to $n\le m$. Evaluating $N-n$
integrals in the points $\{\lambda_{\alpha_{-}}\}$  we arrive at
\eq{resum-x} with $x_k=\xi_k$, and hence $d(x_k)=0$.
Moreover, we can also set  $d(z_k)=0$, since the remaining integrals
surround only the poles at $z_k=\xi_j$ and after evaluation of
these integrals the functions $d(z_k)$ vanish. Finally, using 
the fact that the set $\{\lambda\}$ satisfies the system
\eq{TTMBE} we obtain
\begin{multline}\label{short-exp}
\langle Q^\kappa_{1,m}\rangle=
\sum_{n=0}^{m}
 \sum_{\{\lambda\}=\{\lambda_{\alpha_+}\}\cup\{\lambda_{\alpha_-}\}
 \atop{|\alpha_+|=n}}
 \frac{1}{n!}\oint\limits_{\Gamma\{\xi\}}
 \prod_{j=1}^{n}\frac{dz_j}{2\pi i}
 \cdot
\frac{\prod\limits_{b\in\alpha_+}
U(\lambda_b|\{z\},\{\lambda_{\alpha_-}\},\{\xi\})}
{\prod\limits_{b=1}^n
U(z_b|\{z\},\{\lambda_{\alpha_-}\},\{\xi\})}
\\
\times\det_n \tilde M_\kappa (\{\lambda_{\alpha_+}\},\{z\})
\frac{
\left.\det_N \Omega (\{\lambda\} , \{z\} \cup
\{\lambda_{\alpha_-}\} |\{\lambda\} )\right|_{d(z)=0}}
{\det_N\Omega (\{\lambda\} ,\{\lambda_{\alpha_+}\} \cup \{\lambda_{\alpha_-}\}
| \{\lambda\})},
\end{multline}
where
\be{def-U}
U(\nu|\{z\},\{\lambda_{\alpha_-}\},\{\xi\})= a(\nu)
\prod\limits_{a=1}^n\sinh(z_a-\nu+\eta)
\prod\limits_{a\in\alpha_{-}}\sinh(\lambda_a-\nu+\eta)
\prod_{a=1}^{m}\frac{\sinh(\nu-\xi_a)}{\sinh(\nu-\xi_a+\eta)},
\ee
and
\be{GFtiMjk}
(\tilde M_\kappa)_{jk}(\{\lambda\},\{z\})
=t(z_k,\lambda_j)+\kappa t(\lambda_j,z_k)
\prod_{a=1}^n\frac{\sinh(\lambda_a-\lambda_j+\eta)}
{\sinh(\lambda_j-\lambda_a+\eta)}
\frac{\sinh(\lambda_j-z_a+\eta)}
{\sinh(z_a-\lambda_j+\eta)}.
\ee
Observe that, unlike in the master equation,
the integrand in the r.h.s. of \eq{short-exp} is a polynomial
in $\kappa$, since $\kappa$ enters only the determinant of the matrix
$\tilde M_\kappa$ \eq{GFtiMjk}. Therefore one can set $\kappa$ to be
an arbitrary complex number directly in the integral
representation \eq{short-exp}.

This representation also allows  one  to proceed,
if necessary, to the homogeneous limit simply by setting $\xi_k\to\eta/2$.
Finally, the equation \eq{short-exp} is convenient for taking
the thermodynamic limit $N,M\to\infty$ $N/M=const$. We refer the reader
for the details to the paper \cite{KitMST02a} and here simply recall
the final result for the ground state expectation value
of $\langle Q^\kappa_{1,m}\rangle$ in the thermodynamic limit and
in an external magnetic field in the massive and massless regimes:
\begin{multline}\label{SDE-TD-lim}
\langle Q^\kappa_{1,m}\rangle=\sum_{n=0}^{m}
\frac1{(n!)^2}\oint\limits_{\Gamma\{\xi\}} \prod_{j=1}^n\frac{dz_j}{2\pi i}
\int_C d^n\lambda
\prod_{b=1}^n\prod_{a=1}^{m} \left[\frac{\sinh(\lambda_b-\xi_a)}
{\sinh(z_b-\xi_a)} \frac{\sinh(z_b-\xi_a+\eta)}
{\sinh(\lambda_b-\xi_a+\eta)} \right]\\
\times
\prod_{a=1}^n\prod_{b=1}^n\frac{
\sinh(\lambda_a-z_b+\eta)\sinh(z_b-\lambda_a+\eta)}
{\sinh(\lambda_a-\lambda_b+\eta)\sinh(z_a-z_b+\eta)}
\cdot
{\det}_{n}\Bigl[\tilde M_\kappa(\{\lambda\}|\{z\})\Bigr]
\cdot{\det}_{n}\Bigl[\rho(\lambda_j,z_k)\Bigr].
\end{multline}
Here the function $\rho(\lambda,z)$ is the `inhomogeneous spectral density'
of the ground state, having the support on the contour $C$ 
as defined in \cite{KitMST02a}.


\section{Form factor expansion}\label{f-f-e}

Evaluating the integral \eq{Master-2} by the residues outside the
integration contour we arrive at the expansion over form factors
for $\langle Q^\kappa_{1,m}\rangle$. Recall that due to Lemma
\ref{main-lem}, the only poles outside the contour
$\Gamma\{\xi\}\cup\Gamma\{\lambda\}$ which contribute to the
integral \eq{Master-2}, are the admissible off-diagonal solutions
of \eq{TBE-z}. Hence
\begin{multline}\label{sum-ff}
\langle Q^\kappa_{1,m}\rangle=(-1)^N\sum_{\{\mu\}}
\prod_{a,b=1}^{N}\sinh^2(\lambda_a-\mu_b)
\cdot\prod_{a=1}^m\frac{\tau_\kappa(\xi_a|\{\mu\})}
{\tau(\xi_a|\{\lambda\})}
\\
\times\frac{\det_N
\left(\frac{\partial\tau_\kappa(\lambda_j|\{\mu\})}{\partial \mu_k}\right)
}{\det_N\left(\frac{\partial{\cal Y}_\kappa (\mu_k| \{\mu\})}
{\partial\mu_j}\right)}\cdot
\frac{\det_N
\left(\frac{\partial\tau(\mu_k|\{\lambda\})}{\partial \lambda_j}\right)}
{\det_N \left(\frac{\partial{\cal Y}(\lambda_k|\{\lambda\})}
{\partial\lambda_j}\right)},
\end{multline}
where the sum is taken with respect to all admissible off-diagonal
solutions  $\mu_1,\dots,\mu_N$ of the system \eq{TBE-z}. Due to
the Theorem \ref{Thm-TV} the Jacobian matrix $\partial{\cal
Y}_\kappa (\mu_k| \{\mu\})/\partial\mu_j$ is non-degenerated.

Using the formula \eq{FCdet} we identify
each Jacobian in \eq{sum-ff} with the corresponding scalar products,
leading to
\be{identify}
\langle Q^\kappa_{1,m}\rangle=\sum_{\{\mu\}}
\prod_{a=1}^m\frac{\tau_\kappa(\xi_a|\{\mu\})}
{\tau(\xi_a|\{\lambda\})}\cdot
\frac{\langle\psi (\{\lambda\})|\psi_\kappa(\{\mu\})\rangle}
{\langle\psi_\kappa(\{\mu\})|\psi_\kappa(\{\mu\})\rangle}\cdot
\frac{\langle\psi_\kappa(\{\mu\})|\psi (\{\lambda\})\rangle}
{\langle\psi (\{\lambda\})|\psi (\{\lambda\})\rangle}.
\ee
It remains to use that the state $|\psi_\kappa(\{\mu\})\rangle$
is the eigenstate of ${\cal T}_\kappa(\xi)$ with the eigenvalue
$\tau_\kappa(\xi|\{\mu\})$ and the state
$|\psi (\{\lambda\})\rangle$ is the eigenstate of ${\cal T}(\xi)$
with the eigenvalue $\tau(\xi|\{\lambda\})$.
Thus, the equation \eq{identify} can be written in the form
\be{sum-ff-1}
\langle Q^\kappa_{1,m}\rangle=\sum_{\{\mu\}}
\frac{\langle\psi (\{\lambda\})|
\prod\limits_{b=1}^m{\cal T}_\kappa(\xi_b)|
\psi_\kappa(\{\mu\})\rangle\cdot
\langle\psi_\kappa(\{\mu\})|
\prod\limits_{b=1}^m{\cal T}^{-1}(\xi_b)|
\psi (\{\lambda\})\rangle}
{\langle\psi_\kappa(\{\mu\})|\psi_\kappa(\{\mu\})\rangle\cdot
\langle\psi (\{\lambda\})|\psi (\{\lambda\})\rangle}.
\ee
Observe that we did not use the statement b) of the
Theorem \ref{Thm-TV} on the completeness of the set $|\psi_\kappa(\{\mu\})
\rangle$. The sum over eigenstates of ${\cal T}_\kappa$ appears automatically as
the result of the evaluation of the multiple integral \eq{Master-2} by
the residues outside the integration contour.

Taking the second lattice derivative of \eq{sum-ff} and then
differentiating twice with respect to $\kappa$ at $\kappa=1$
(see Appendix \ref{FF-sigma}) we
obtain the form factor type expansion directly for the correlation function
of the third components of the spin
\be{ff-expan-s}
\langle\sigma_{1}^z\sigma_{m+1}^z\rangle=
\langle\sigma_{1}^z\rangle\cdot\langle\sigma_{m+1}^z\rangle\\
+\sum_{\{\mu\}\ne\{\lambda\}}
\frac{\langle\psi (\{\lambda\})|\sigma_{1}^z|
\psi(\{\mu\})\rangle\cdot
\langle\psi(\{\mu\})|\sigma_{m+1}^z|
\psi (\{\lambda\})\rangle}
{\langle\psi(\{\mu\})|\psi(\{\mu\})\rangle\cdot
\langle\psi (\{\lambda\})|\psi (\{\lambda\})\rangle},
\ee
where the form factors of $\sigma^z$ are given by \eq{ff-sigma}. Observe that here the
summation is taken with respect to the eigenstates of the operator
${\cal T}$, but not ${\cal T}_\kappa$. In other words, in the homogeneous
limit, this is the sum over the excited states of the Hamiltonian.

Of course, the equation \eq{ff-expan-s} might be obtained by means
of inserting the complete set of the eigenstates of the
Hamiltonian between local spin operators. Such a representation,
however, would be quite formal without a precise description of the
set of the states on which the summation has to be performed. 
On the one hand it is clear that the sum
should be taken with respect to solutions of the system
\eq{TTMBE}, but on the other hand it is also clear that not any
solution of this system corresponds to an eigenstate of the
Hamiltonian. Note that one can not simply restrict the sum in
\eq{ff-expan-s} to the admissible off-diagonal solutions,
since in the homogeneous limit certain unadmissible solutions of
the system \eq{TTMBE} give rise to the basis of the eigenstates.

The master equation approach gives the way to overcome these
difficulties. Namely, if we use the approach \eq{int-ff}, we can
insert the set of the eigenstates of the twisted transfer matrix
between local operators. Then the  summation in
\eq{ff-expan-s} is restricted to the admissible off-diagonal solutions
of the system \eq{TTMBE_Y}. The unadmissible solutions
corresponding to the eigenstates of the Hamiltonian appear in this
case only as the limit of admissible solutions at $\kappa\to1$.
Since for $|\kappa|$  small enough all admissible solutions are in
the vicinities of the points $\{\xi-\eta\}$, the sum over excited
states can be written as a multiple integral with respect to a
contour surrounding these points. This allows one to obtain the master
type equation directly for the spin--spin correlation functions, 
starting form their form factor expansions.


\section{Conclusion}

The main result of this paper is the master equation
\eq{Master-2}. We have shown that this equation draws a link
between multiple integral representations and form factor expansions
for the correlation functions.

The master equation sheds a new light on the role of the auxiliary
contour integrals in the representations for the correlation
functions derived in \cite{KitMST02a}. Originally they appeared
as a result of a re-summation of the elementary blocks obtained
in \cite{JimMMN92,JimM96,KitMT00}. In the framework
of the master equation approach the same auxiliary contour
integrals are equivalent to the sum over excited states.

Using the master equation method one can derive multiple integral representations
for other correlation functions of the $XXZ$ model, including many-point
correlators. Indeed, an arbitrary correlation function in the finite chain
can be reduced to the multiple sum over complete set of the eigenstates of
the twisted transfer matrix. On the other hand the explicit formulas for
the form factors of the local spin operators obtained in \cite{KitMT99}
can be easily generalized for the case when one or both states are eigenstates
of the twisted transfer matrix. Presenting this sum as a contour
integral of the type \eq{Master-2} and evaluating this integral
by the residues inside this contour we can express 
an arbitrary correlation function of the $XXZ$ model
as a multiple integral of the form \eq{SDE-TD-lim}.

Moreover, we conjecture that representations similar to \eq{Master-2} should
exist for the correlation functions of other models solvable by the
algebraic Bethe ansatz. This conjecture is based on the fact that the integrand of
\eq{Master-2} depends mostly on the eigenvalues of the twisted transfer
matrix, which is a typical object for the
algebraic Bethe ansatz. As we have mentioned already in Introduction, it
would be very desirable to obtain an analog of this master equation in the
case of the field theory models. For these models such an integral representation could give
an analytic link between short distance and long distance expansions of
the correlation functions.

Finally, the master equation method opens a way to obtain multiple integral
representations for time-dependent correlation functions. This
subject will be considered in our forthcoming publication \cite{KitMST04b}. Here
we only announce a generalization of the representation \eq{SDE-TD-lim} for
the time-dependent case in the massive and massless regimes

\be{New-funct}
\langle\sigma_{1}^z(0)\sigma_{m+1}^z(t)\rangle=
2\langle\sigma_{1}^z(0)\rangle-1+2D^2_m
\left.\frac{\partial^2}{\partial\kappa^2}{\cal Q}_\kappa(m,t)\right|_{\kappa=1},
\ee
where $D^2_m$ means the second lattice derivative and
\ba{answer-time}
&&{\dis\hspace{-7mm}
{\cal Q}_\kappa(m,t)=
\sum_{n=0}^{\infty}\frac{1}{(n!)^2}\int\limits_C d^n\lambda
\oint\limits_{\Gamma\{\pm\frac\eta2\}}\prod_{j=1}^{n}
\frac{dz_j}{2\pi i}\cdot \prod_{a,b=1}^n\frac{
\sinh(\lambda_a-z_b+\eta)\sinh(z_b-\lambda_a+\eta)}
{\sinh(\lambda_a-\lambda_b+\eta)\sinh(z_a-z_b+\eta)}
}\nona{40}
&&{\dis\hspace{-12mm}
\times
\prod_{b=1}^ne^{it(E(z_b)-E(\lambda_b))+im(p(z_b)-p(\lambda_b))}\;
\det_n\tilde M_\kappa(\{\lambda\}|\{z\})
\cdot\det_n[{\cal R}^{\kappa}_n(\lambda_j,z_k|\{\lambda\},\{z\})].}
\ea
Here $\tilde M_\kappa$ is given by \eq{GFtiMjk}. The functions
$E(\lambda)$ and $p(\lambda)$ are the bare one-particle energy
and momentum
\be{E}
E(\lambda)=\frac{2\sinh^2\eta}
{\sinh(\lambda+\frac{\eta}{2})\sinh(\lambda-\frac{\eta}{2})},\qquad
p(\lambda)=i\log\left(\frac{\sinh(\lambda-\frac{\eta}{2})}
{\sinh(\lambda+\frac{\eta}{2})}\right).
\ee
The contour $\Gamma\{\pm\eta/2\}$ surrounds the points
$\pm\eta/2$ and does not contain any other singularities of the integrand.
The function ${\cal R}_n^\kappa(\lambda,z|\{\lambda\},\{z\})$, as
a function of $z$ and other arguments fixed,
has a cut between the points $z=\eta/2$ and $z=-\eta/2$. Therefore
it is defined differently in the vicinities of these points
\be{calR}
{\cal R}^{\kappa}_n(\lambda,z|\{\lambda\},\{z\})=\left\{
\begin{array}{l}
\rho(\lambda,z),\qquad z\sim\eta/2;\\
-\kappa^{-1}\rho(\lambda,z+\eta)\cdot
\prod_{b=1}^{n}\frac{\sinh(z-\lambda_b+\eta)\sinh(z_b-z+\eta)}
{\sinh(\lambda_b-z+\eta)\sinh(z-z_b+\eta)}, \qquad
z\sim-\eta/2.
\end{array}\right.,
\ee
where $\rho(\lambda,z)$ is the inhomogeneous spectral density
of the ground state. 

Observe that the time and the distance dependence of the generating
function \eq{answer-time} is associated to the bare energy $E(\lambda)$
and momentum $p(\lambda)$, making the equation \eq{answer-time}
very suggestive. In the limit $t=0$ the integrand in \eq{answer-time}
is a holomorphic function in the vicinity of $-\eta/2$, therefore
the integrals over $\Gamma\{-\eta/2\}$ vanish, and we arrive
at the time-independent representation \eq{SDE-TD-lim}.


\section*{Acknowledgments}
J. M. M., N. S. and V. T. are supported by CNRS. N. K., J. M. M.,
V. T. are supported by the European network
EUCLID-HPRNC-CT-2002-00325.  J. M. M. and N.S. are supported
by INTAS-03-51-3350. N.S. is supported
by the French-Russian Exchange Program, the Program of RAS
Mathematical Methods of the Nonlinear Dynamics, 
RFBR-02-01-00484, Scientific Schools 2052.2003.1. 
N. K, N. S. and V. T. would like to thank the Theoretical Physics group 
of the Laboratory of Physics at ENS Lyon for hospitality, which makes
this collaboration possible.


\appendix

\section{Solutions of the twisted Bethe equations}\label{Sol-TBE}

In this Appendix we prove Lemmas \ref{main-lem-0},
\ref{main-lem} using the results of \cite{TarV96}. 
Consider the system of equations \eq{TBE-z}
\be{Polyn-Bethe}
{\cal Y}_\kappa(z_j|\{z\})=0,\qquad j=1,\dots,N.
\ee
Without loss of generality we can identify two solutions of this
system $\{z\}$ and $\{z'\}$ if they are equal modulo $i\pi$.

It was proved in \cite{TarV96} that

\begin{thm}\label{Thm-TV}\cite{TarV96}
Let $\kappa$ and inhomogeneities $\{\xi\}$ be generic. Then

a) All admissible off-diagonal solutions of the system
\eq{Polyn-Bethe} are non-degenerated.

b) The set of the states \eq{ABA-TES} corresponding to the admissible
off-diagonal solutions form a basis in the subspace ${\cal
H}^{(M/2-N)}$.
\end{thm}

The strategy of \cite{TarV96} was to consider the limit of the system
\eq{Polyn-Bethe} at $\kappa\to 0$ and then to deform the solutions
$\{z(0)\}$ to the case $\kappa\ne0$. Then the statement a) follows from the
implicit function theorem.  Indeed, the solutions of the system
\eq{Polyn-Bethe} at $\kappa=0$ are evident. In particular all
admissible solutions have form $z_j(0)=\xi_{p_j}-\eta$. One can
easily check that the Jacobian matrix $\partial{\cal
Y}_\kappa(z_j|\{z\})/\partial z_k$ at $\kappa=0$ for admissible
solutions is a diagonal matrix with non vanishing entries. Hence, 
in this case the solution $\{z(\kappa)\}$ is a holomorphic deformation of the
solution $\{z(0)\}$, and therefore for 
$|\kappa|$ small enough all admissible  solutions are in the vicinities of
$\{\xi-\eta\}$. On the other hand it is clear that any admissible solution at
$\kappa\ne1$  is separated from the admissible solutions at
$\kappa=1$. Thus, admissible solutions are separated from the
points $\{\xi\}$ and $\{\lambda\}$. This proves  Lemma
\ref{main-lem-0}.

In order to prove Lemma \ref{main-lem} we use the properties
of the matrix $\Omega_\kappa(\{\lambda\},\{\mu\}|\{\nu\})$ \eq{matH}
formulated in the following two lemmas.
Let the parameters $\lambda_1,\dots,\lambda_n$
and  $\mu_1,\dots,\mu_n$ be generic complex numbers, $\kappa$ be
arbitrary complex.

\begin{lemma}\label{family}
Suppose there exist $\mu_a,\mu_b\subset\{\mu\}$, such that $\mu_a=\xi_p, \quad
\mu_b=\xi_p-\eta$, where
$\xi_p$ is one of the inhomogeneity parameters. Then
$\det_n \Omega_\kappa(\{\lambda\},\{\mu\}|\{\lambda\})=0$.
\end{lemma}
{\sl Proof.} We have $d(\mu_a)=a(\mu_b)=0$ and $t(\mu_b,\lambda_j)
=t(\lambda_j,\mu_a)$. Thus, the
columns $(\Omega_\kappa)_{ja}$ and $(\Omega_\kappa)_{jb}$
are proportional to each other, hence ${\det}_n \Omega_{\kappa}=0$.  
$\Box$

\begin{lemma}\label{string-t}
Let $q^2=e^{2\eta}$ be a root of unity, i.e. $\eta=i\pi Q/P$, where $Q<P$ are 
positive integers. Suppose there exists $\{\lambda_{a_1},\dots \lambda_{a_P}\}
\subset\{\lambda\}$, such that
\be{cond}
\sinh(\lambda_{a_{j+1}}-\lambda_{a_{j}}+\eta)=0,
\qquad\mbox{with}\qquad \lambda_{a_{P+1}}\equiv\lambda_{a_{1}}.
\ee
Then $\det_n \Omega_\kappa(\{\lambda\},\{\mu\}|\{\lambda\})=0$.
\end{lemma}
{\sl Proof.} It is easy to see that 
\be{easy-sum}
\sum_{j=1}^Pt(\mu,\lambda_{a_j})=\sum_{j=1}^P t(\lambda_{a_j},\mu)=0,
\ee
since both these sums are $i\pi$-periodical holomorphic functions of $\mu$
vanishing at $\mu\to\pm\infty$. Hence the matrix
$\Omega_\kappa(\{\lambda\},\{\mu\},\{\lambda\})$ contains linearly
depended lines and its determinant vanishes.$\Box$

Note that if the set $\{\lambda\}$ contains several  subsets of the type
\eq{cond}, then the order of the zero of 
$\det_n \Omega_\kappa(\{\lambda\},\{\mu\}|\{\lambda\})$ is not less than
the number of such subsets.

If $q^2$ is not root of unity, then unadmissible solutions of the system 
\eq{Polyn-Bethe} contain a pair $z_a=\xi_p$ and $z_b=\xi_p-\eta$.
Off-diagonal unadmissible solutions are non-degenerated
at $\kappa=0$ \cite{TarV96}. Hence, due to Lemma
\ref{family} they do not give a contribution to the integrals
\eq{Master-1}, \eq{master1}. The diagonal solutions 
$\{z(\kappa)\}$ of \eq{Polyn-Bethe}
obtained as a deformation of diagonal solutions
$\{z(0)\}$ at $\kappa=0$ preserve their  multiplicity
\cite{TarV96}. Hence, they also do not give any contribution to
the integrals \eq{Master-1}, \eq{master1} due to vanishing of 
$\det_N \Omega_\kappa$ and $\det_N \Omega$
as soon as two or more $z$ coincide.

If $q^2$ is a root of unity, then unadmissible solutions may contain 
`strings' $\{z_{a_j}\}$ satisfying the condition \eq{cond}. Such solutions
are not isolated since the value of $z_{a_1}$ is not fixed. We have seen 
however that due to Lemma \ref{string-t} the determinant
of the matrix $\Omega_\kappa(\{z\},\{\lambda\}|\{z\})$ vanishes exactly
on these string type solutions. Hence, these unadmissible
solutions of the system \eq{Polyn-Bethe} do not contribute to
the integrals \eq{Master-1}, \eq{master1}.

Thus, the only solutions of the system \eq{Polyn-Bethe} which can
give non-vanishing contribution to the integrals \eq{Master-1}, \eq{master1} 
are admissible off-diagonal solutions, which are in the vicinities of
$\{\xi-\eta\}$ for $|\kappa|$ small enough.
This proves Lemma \ref{main-lem}.


\section{The  $\sigma^z$ form factor}\label{FF-sigma}

The explicit formulas for form factors of local spin operators in the finite
$XXZ$ chain were obtained in \cite{KitMT99}. Here we 
propose slightly modified method
to derive the form factor of the operator $\sigma^z$.

Consider a matrix element of the operator $Q^\kappa_{1,m}$ between an eigenstate
$|\psi(\{\lambda\})\rangle$ of the transfer matrix ${\cal T}(\nu)$ (for example,
ground state) and an eigenstate $\langle\psi_\kappa(\{\mu\})|$ of the operator
${\cal T}_\kappa(\nu)$. Using \eq{GFexpbQ} we immediately
obtain
\be{matr-el}
\langle\psi_\kappa(\{\mu(\kappa)\})|Q^\kappa_{1,m}
|\psi(\{\lambda\})\rangle=
\prod_{a=1}^m\frac{\tau_\kappa(\xi_a|\{\mu(\kappa)\})}
{\tau(\xi_a|\{\lambda\})}\cdot
\langle\psi_\kappa(\{\mu(\kappa)\})|\psi (\{\lambda\})\rangle.
\ee
Here we have indicated explicitly that the parameters $\{\mu\}$ depend
on $\kappa$, for they are solutions of the system \eq{TTMBE_Y}.
On the other hand it is clear that
\be{other-hand}
\left.\frac\partial{\partial\kappa}
\langle\psi_\kappa(\{\mu(\kappa)\})|[Q^\kappa_{1,m+1}
-Q^\kappa_{1,m}]|\psi(\{\lambda\})\rangle\right|_{\kappa=1}=
\frac12\langle\psi(\{\mu(1)\})|(1-\sigma_{m+1}^z)|\psi (\{\lambda\})\rangle.
\ee
Thus, we have
\begin{multline}\label{ev-res}
\frac12\langle\psi(\{\mu(1)\})|(1-\sigma_{m+1}^z)|\psi (\{\lambda\})\rangle
=\frac\partial{\partial\kappa}
\prod_{a=1}^m\frac{\tau_\kappa(\xi_a|\{\mu(\kappa)\})}
{\tau(\xi_a|\{\lambda\})}\cdot\left(
\frac{\tau_\kappa(\xi_{m+1}|\{\mu(\kappa)\})}
{\tau(\xi_{m+1}|\{\lambda\})}-1\right)
\\
\times
\frac{\prod_{a=1}^{N} d(\mu_a(\kappa))}
{\prod\limits_{a>b}^N\sinh(\lambda_a-\lambda_b)\sinh(\mu_b(\kappa)
-\mu_a(\kappa))}
\cdot \left.\det_N \Omega_\kappa(\{\mu(\kappa)\},\{\lambda\}|\{\mu(\kappa)\})
\right|_{\kappa=1}.
\end{multline}

In order to evaluate explicitly the derivative over $\kappa$ in \eq{ev-res}
one should distinguish two cases: $\{\mu(1)\}=\{\lambda\}$ and
$\{\mu(1)\}\ne\{\lambda\}$. In the first case $\tau_\kappa(\xi_{m+1}|\{\mu(\kappa)\})
\to\tau(\xi_{m+1}|\{\lambda\})$ as $\kappa\to1$, therefore
\begin{align}\label{case-1}
\frac{\langle\psi(\{\lambda\})|(1-\sigma_{m+1}^z)|\psi (\{\lambda\})\rangle}
{\langle\psi(\{\lambda\})|\psi (\{\lambda\})\rangle} &=
\left.2\frac\partial{\partial\kappa}
\left(\frac{\tau_\kappa(\xi_{m+1}|\{\mu(\kappa)\})}
{\tau(\xi_{m+1}|\{\lambda\})}-1\right)\right|_{\kappa=1}
\\
&=-2\sum_{k=1}^N \left.\frac{d\mu_k(\kappa)}{d\kappa}
\right|_{\kappa=1}\cdot t(\lambda_k,\xi_{m+1}).
\end{align}
The derivatives $d\mu_k(\kappa)/d\kappa$ can be found from \eq{TBE-l-1} via
\be{Impl-dif}
\sum_{k=1}^N\frac{\partial {\cal Y}_\kappa(\mu_j|\{\mu\})}{\partial\mu_k}
\cdot \left.\frac{d\mu_k(\kappa)}{d\kappa}
\right|_{\kappa=1}+d(\mu_j)\prod_{a=1}^N\sinh(\mu_a-\mu_j-\eta)=0.
\ee

In the second case we can compute explicitly the derivative of  $\det \Omega_\kappa$.
Indeed, consider the $N$-dimensional vector-column $v$ with the
components
\be{PSNev}
v_k=\prod_{a=1}^N\sinh(\mu_k(\kappa)-\lambda_a)
\prod_{a=1\atop{a\ne k}}^N\sinh^{-1}(\mu_k(\kappa)-\mu_a(\kappa)).
\ee
If $\{\mu(1)\}\ne\{\lambda\}$, then this vector at $\kappa=1$ has at
least one non-zero component, say $v_N\ne 0$. Then multiplying
the $k$-th column of $(\Omega_\kappa)_{jk}$ by $v_k/v_N$ and adding
the first $(N-1)$ columns to the last one, we obtain
\be{add-column}
(\Omega_\kappa)_{jN}+\sum_{k=1}^{N-1}\frac{v_k}{v_N}(\Omega_\kappa)_{jk}
=\frac1{v_N}{\cal Y}_\kappa(\lambda_j|\{\lambda\})=
\frac{1-\kappa}{v_N}\cdot a(\lambda_j)\prod_{a=1}^N
\sinh(\lambda_a-\lambda_j+\eta).
\ee
(see Appendix B of \cite{KitMST02a} for the proof). Thus, the last column
is proportional to $1-\kappa$, hence, taking the derivative over $\kappa$ of
$\det \Omega_\kappa$ one has to differentiate only this column. This gives
\be{deriv-H}
\frac\partial{\partial\kappa}
\left.\det_N \Omega_\kappa(\{\mu(\kappa)\},\{\lambda\}|\{\mu(\kappa)\})
\right|_{\kappa=1}
=\det_N \tilde \Omega(\{\mu(1)\},\{\lambda\}|\{\mu(1)\}),
\ee
where
\be{tilde-H}
(\tilde \Omega)_{jk}(\{\mu\},\{\lambda\}|\{\mu\})=
\left\{\begin{array}{l}
\left.(\Omega_\kappa)_{jk}(\{\mu\},\{\lambda\}|\{\mu\})
\right|_{\kappa=1},\qquad
k=1,\dots,N-1,\non
-\frac1{v_N}\cdot a(\lambda_j)\prod_{a=1}^N
\sinh(\lambda_a-\lambda_j+\eta),\qquad k=N.
\end{array}\right.
\ee
Thus, for $\{\mu\}\ne\{\lambda\}$ we obtain
\begin{multline}\label{ff-sigma}
\langle\psi(\{\mu\})|\sigma_{m+1}^z|\psi (\{\lambda\})\rangle
=2\prod_{a=1}^m\frac{\tau_\kappa(\xi_a|\{\mu\})}
{\tau(\xi_a|\{\lambda\})}\cdot\left(
1-\frac{\tau_\kappa(\xi_{m+1}|\{\mu\})}
{\tau(\xi_{m+1}|\{\lambda\})}\right)
\\
\times
\frac{\prod_{a=1}^{N} d(\mu_a)}
{\prod\limits_{a>b}^N\sinh(\lambda_a-\lambda_b)\sinh(\mu_b-\mu_a)}
\cdot \det_N \tilde \Omega(\{\mu\},\{\lambda\}|\{\mu\}).
\end{multline}
%




\begin{thebibliography}{99}
%
\bibitem{Bax82L} R. J. Baxter, {\it Exactly solved models
in statistical mechanics}, Academic Press,
London--New-York, 1982.
%
\bibitem{Fad84} L.D. Faddeev, Les Houches 1982, {\it Recent advances in field
theory and statistical mechanics}, edited by J.B. Zuber and R. Stora,
Elsevier Science Publ., 1984, 561.
%
\bibitem{GauL83} M. Gaudin,
{\it La Fonction d'Onde de Bethe}, Paris: Masson, 1983.
%
\bibitem{BogIK93L}V. E. Korepin, N. M. Bogoliubov,
A. G. Izergin,
{\it Quantum Inverse Scattering Method and Correlation
Functions}, Cambridge University Press, 1993.
%
\bibitem{Ons44} L. Onsager, Phys. Rev. {\bf 65} (1944) 117.
%
\bibitem{LieSM61}  E. Lieb, T. Shultz and D. Mattis,
Ann. Phys., {\bf 16} (1961)  407.
%
\bibitem{Mcc68}  B. M. McCoy, Phys. Rev.,
{\bf 173} (1968)  531.
%
\bibitem{WuMTB76}
T. T. Wu, B. M; McCoy, C. A. Tracy, E. Barouch,
Phys. Rev. B {\bf
13} (1976) 316.
%
\bibitem{MccTW77}  B. M. McCoy, C. A. Tracy and T. T, Wu,
Phys. Rev. Lett., {\bf 38} (1977)  793.
%
\bibitem{SatMJ78} M. Sato, T. Miwa, M. Jimbo,
Publ. Res. Int. Math. Sci. {\bf 14} (1978) 223; {\bf 15} (1979)
201, 577, 871; {\bf 16} (1980) 531.
%
\bibitem{BelPZ84} A.A. Belavin, A.M. Polyakov, A.B. Zamolodchikov,
Nucl.Phys. B {\bf 241} (1984) 333.
%
\bibitem{KarW78} M.Karowski and P Weisz, Nucl.Phys. B {\bf 139} (1978) 455.
%
\bibitem{Smi92L} F.A. Smirnov, {\it Form factors in completely integrable models
of quantum field theory}. Advanced series in mathematical physics,
14, World Scientific (1992).
%
\bibitem{BabFKZ99}
H. Babudjan, A. Fring, M.Karowski, A. Zapletal,
Nucl.Phys. B {\bf 538} (1999) 535.
%
\bibitem{Hei28} W. Heisenberg, Zeitschrift f\"ur
Physik, {\bf 49} (1928)  619.
%
\bibitem{Bet31} H. Bethe, Zeitschrift f\"ur
Physik, {\bf 71} (1931)  205.
%
\bibitem{Orb58} R. Orbach, Phys. Rev.,
{\bf 112} (1958)  309.
%
\bibitem{Wal59} L. R. Walker, Phys. Rev.,
{\bf 116} (1959)  1089.
%
\bibitem{YanY66a} C. N, Yang and C. P. Yang, Phys. Rev.,
{\bf 150} (1966)  321, 327.
%
\bibitem{FadST79} L. D. Faddeev, E. K. Sklyanin and L. A. Takhtajan,
Theor. Math. Phys. {\bf 40} (1980) 688.
%
\bibitem{TakF79} L. A. Takhtajan and  L. D. Faddeev,
Russ. Math. Surveys. {\bf 34} (1979) 11.
%
\bibitem{JimMMN92}  M. Jimbo, K.~Miki, T. Miwa and  A.~Nakayashiki,
Phys. Lett. A  {\bf 168} (1992) 256.
%
\bibitem{JimM96}  M. Jimbo and T. Miwa,
Journ. Phys. A: Math. Gen., {\bf 29} (1996) 2923.
%
\bibitem{JimML95}  M. Jimbo and T. Miwa,
Algebraic analysis of solvable lattice models (AMS, 1995).
%
\bibitem{KitMT99} N. Kitanine, J. M. Maillet and V. Terras,
Nucl. Phys. B, {\bf 554} [FS] (1999) 647, math-ph/9807020.
%
\bibitem{KitMT00} N. Kitanine, J. M. Maillet and V. Terras,
Nucl. Phys. B, {\bf 567} [FS] 554. (2000), math-ph/9907019
%
\bibitem{MaiT99} J. M. Maillet and V. Terras,
Nucl. Phys. B 575 (2000) 627, hep-th/9911030.
%
\bibitem{IzeKMT99} A. G. Izergin, N. Kitanine, J. M. Maillet and V. Terras,
Nucl. Phys. B, {\bf 554} [FS] (1999) 679.
%
\bibitem{BosKS02} H. Boos, V. Korepin, F. Smirnov,
J.Phys. A {\bf 37} (2004) 323.
%
\bibitem{BosJMST04} H. Boos, M. Jimbo, T. Miwa, F. Smirnov,
Y. Takeyama, {\it A recursion formula for the correlation
functions of an inhomogeneous $XXX$ model}, hep-th/0405044.
%
\bibitem{KitMST02a}
N. Kitanine, J. M. Maillet, N. A. Slavnov, V. Terras,
Nucl. Phys. B {\bf 641} [FS] (2002) 487; hep-th/0201045.
%
\bibitem{KitMST02b}
N. Kitanine, J. M. Maillet, N. A. Slavnov, V. Terras, Nucl. Phys.
B {\bf 642} [FS] (2002) 433; hep-th/0203169.
%
\bibitem{GohKS04} F. G\"ohmann, A. Klumper, A. Seel, {\it
Integral representations for the correlation functions of the $XXZ$
chain at finite temperature}, hep-th/0405089.
%
\bibitem{Mik94}
K. Miki, Phys. Lett. A {\bf 186} (1994) 217.
%
\bibitem{JimKMQ94} M. Jimbo, T. Kojima, T. Miwa, Y. H. Quano, J.
Phys. A {\bf 27} (1994) 3267.
%
\bibitem{KojMQ95} T. Kojima, F. Miki, Y. H. Quano, J. Phys. A {\bf 28} (1995) 3479.
%
\bibitem{Qua98} Y. H. Quano, J. Phys. A {\bf 31} (1998) 1791.
%
\bibitem{BabB92}
O. Babelon, D. Bernard, Phys.Lett. B {\bf 288} (1992) 113.
%
\bibitem{ColIKT93} F. Colomo, A. G. Izergin and V. E. Korepin
and V. Tognetti, Theor. Math. Phys. {\bf 94} (1993) 11.
%
\bibitem{Zam91} Al. B. Zamolodchikov, Nucl. Phys. B {\bf 348}
(1991) 619.
%
\bibitem{Zam95} Al. B. Zamolodchikov, Int. J. Mod. Phys. A {\bf
10} (1995) 1125.
%
\bibitem{BabK03}
H. Babudjan, M. Karowski,
{\it Towards the
construction of Wightman functions of integrable quantum field
theories}, hep-th/0301088.
%
\bibitem{BelBLPZ04}
A. A. Belavin, A. V. Belavin, A. V. Litvinov, Y. P. Pugai, Al. B.
Zamolodchikov
Nucl. Phys. B {\bf 676} (2004) 587.
%
\bibitem{TarV96}
V. Tarasov and A. Varchenko, Int.Math.Res.Notices, {\bf 13} (1996)
%
\bibitem{Sla89} N. A. Slavnov,
Theor. Math. Phis. {\bf 79} (1989) 502.
%
\bibitem{IzeK85}A. G. Izergin and V. E. Korepin,
Commun. Math. Phys. {\bf 99}  (1985) 271.
%
\bibitem{GauMW81}  M. Gaudin, B. M. McCoy and T.T. Wu, Phys. Rev. D,
{\bf 23} (1981)  417.
%
\bibitem{Kor82} V.~E.~Korepin,
Commun. Math. Phys. {\bf 86}  (1982) 391.
%
\bibitem{KitMST04b}N. Kitanine, J. M. Maillet, N. A. Slavnov, V. Terras,
{\it Dynamical correlation functions of the $XXZ$ spin-$1/2$
chain}, to be published.

\end{thebibliography}
\end{document}